\documentclass[aps,twocolumn,prb]{revtex4}
\usepackage[OT4]{fontenc}
\usepackage[dvips]{graphicx}
\usepackage{amsmath}
\usepackage{bm}
\usepackage{psfrag}
\usepackage{amssymb}

\usepackage{hyperref}
\usepackage{appendix} 
\usepackage[utf8]{inputenc}
\usepackage[dvipsnames]{color}
\usepackage{fancyhdr}
\usepackage{latexsym}
\usepackage{tgpagella}

\newcommand{\bee}{\begin{eqnarray}}
\newcommand{\eee}{\end{eqnarray}}

\begin{document}

\bibliographystyle{plain}

\title{{\bf Break-up of suspension drops settling under gravity in a viscous fluid close to a vertical wall}}
\author{Anna My\l yk$^*$, Walter Meile$^\dagger$, G\"{u}nter Brenn$^\dagger$ and Maria L. Ekiel-Je\.{z}ewska$^*$\\
$^*$Institute of Fundamental Technological Research, Polish Academy of Sciences, Pawi\'{n}skiego 5B, 02-106 Warsaw, Poland\\
$^\dagger$Institute of Fluid Mechanics and Heat Transfer, University of Technology, Inffeldgasse 25/F, 8010 Graz, Austria}
\date{\today}
\begin{abstract}
The evolution of suspension drops sedimenting under gravity in a viscous fluid close to a vertical wall was studied experimentally and numerically with the use of the point-force model, in the Stokes flow regime. The fluid inside and outside the drop was identical. 
The initial distribution of the suspended solid heavy particles was uniform inside a spherical volume. In the experiments and in the simulations, the suspension drops evolved qualitatively in the same way as in an unbounded fluid. However, it was observed, both experimentally and numerically that, on the average, the destabilization time $T$ and the distance $L$ traveled by the drop until break-up were smaller for a closer distance $h$ of the drop center from the wall, with approximately linear dependence of $T$ and $L$ on $D/h$, for $h$ larger or comparable to the drop diameter $D$. Destabilization times and lengths of individual drops with different random configurations of the particles were shown to differ significantly from each other, owing to the chaotic nature of the particle dynamics. 
\end{abstract}

\maketitle

\section{Introduction}

Motion, deformation and break-up of suspension drops settling under gravity in a viscous fluid has been observed and studied for many years by different researchers.\cite{art66,art10,art21,art26,art9,art28,art12} The most interesting feature of such a process is that a suspension drop settling under the influence of gravity in an unbounded viscous fluid remains a cohesive entity for a long time, even in the absence of a surface tension at the drop surface, and without any attractive direct interactions between the suspended particles.\cite{art9,art21} 
This
property is interesting as a challenging basic theoretical problem, as well as a typical process observed in many practical contexts, such as deposition of drugs in the human airways,\cite{cigarette} rising of mantle plumes,\cite{lister} or interaction of melamine formaldehyde with a complex plasma.\cite{schwabe} 

There is a lot of interest in studying deformation and destabilization of suspension micro-droplets, especially such which consist of a cloud of non-Brownian heavy particles which are denser than the fluid and are separated from each other by a fluid identical to the host fluid outside the drop.\cite{art10,art21,art26,art9,art28} For such systems (on which we focus in this paper), the fluid inertia is negligible, and the fluid flow can be described by the Stokes equations.\cite{Kim-Karrila:1991,Happel-Brenner:1986} Based on the similarity law,\cite{Batchelor} the dynamics of such micro-systems is identical to the dynamics of geometrically similar, but larger and faster objects which move in a fluid of a higher viscosity, provided that the Reynolds number $Re$ for both systems is the same. Suspension drops settling in very viscous fluids at $Re\!<<\!1$ have been studied experimentally by standard video tracking.\cite{art9,art21}
 
The typical evolution pattern of a single suspension drop settling far from walls or interfaces is the following. Initially, the injected drop tends to become spherical, loosing a large number of particles, which move more slowly and therefore form a thin tail behind (above) the drop. The particles suspended in the drop recirculate, settling down faster than the center of mass in the inner part of the drop, and then moving to the outer part of the drop and settling there more slowly than the center of mass, then coming back to the inner part, and so on. At the top of the drop, some of the particles separate out; they move more slowly than the drop and form a thin tail above it. The drop slowly expands sideways and becomes more and more flat. A hole inside is formed, and the resulting torus grows horizontally and decreases its height. Then, the drop suddenly breaks into two (or sometimes more) smaller droplets, which repeat the same evolution pattern. 

This scenario is qualitatively reproduced with the use of various numerical techniques for solving the low-Reynolds-number hydrodynamic equations: the simple point-particle model \cite{art9,art21} or the Rotne-Prager approximation \cite{art28} (for $Re=0$), and the lattice-Boltzmann method,\cite{Tony} spectral methods \cite{Bosse} or Oseen interactions \cite{art12} (for a finite Reynolds number).

The goal of the present paper is to investigate experimentally and theoretically the influence of a solid vertical wall on the evolution, deformation and destabilization of the suspension drops described above, with the emphasis on the destabilization time and the distance a suspension drop travels before breaking (destabilization length). The motivation comes from the previous work,\cite{art67} where simple experiments confirmed that the presence of one or two parallel vertical walls reduces the destabilization length. The effect of the wall on the destabilization of suspension drops is important for practical applications. 

The point-particle model in an unbounded fluid was successfully applied in the literature to determine basic features of a suspension drop evolution.\cite{art10,art21,art26,art9} Therefore, in this work, we use as the theoretical framework the point-particle model close to a plane solid wall, parallel to the direction of gravity. This model has been recently developed and tested elsewhere.\cite{Mylyk_Ekiel-Jezewska:2011} To perform the measurements, we use the experimental set-up applied in a number of previous publications.\cite{art21,art18} 

In Sec.~\ref{2}, we describe our experimental setup. In Sec.~\ref{3}, it is used to generate data on the evolution pattern of suspension drops and their destabilization times and lengths. In Sec.~\ref{3.5}, we introduce a simplified geometrical model of the internal structure of the suspension drop, to allow for efficient numerical modeling.
In Sec.~\ref{4} we introduce the point-particle model close to a wall, present numerical simulations of suspension drops, and determine destabilization times and lengths. Secs.~\ref{compare} and \ref{6} are devoted to comparison between the measurements and the theory, discussion and conclusions.

\section{Experimental set-up}\label{2}
\begin{figure}[h]  
\begin{center}
{\includegraphics[width=8.6cm]{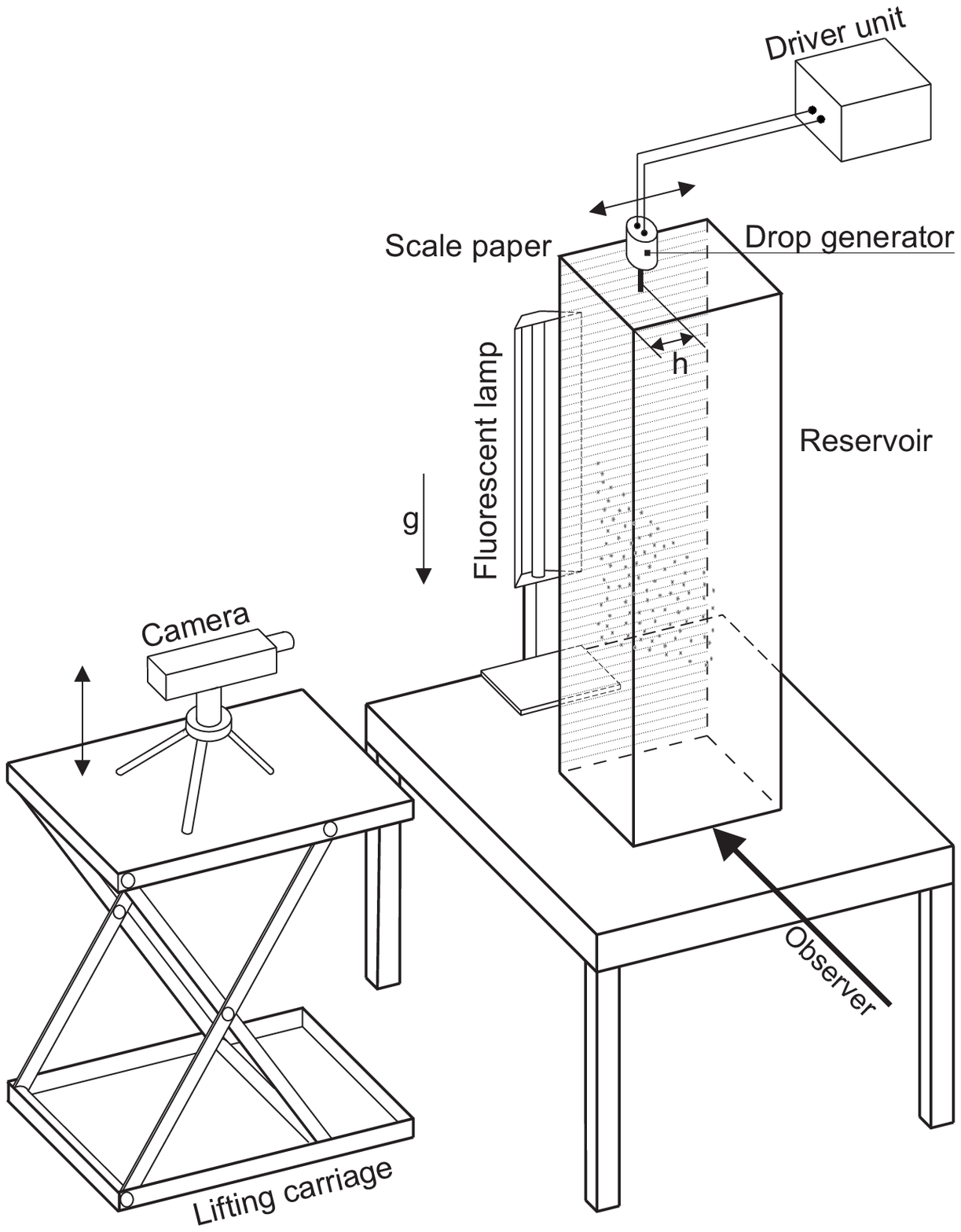}}
{\includegraphics[width=8.6cm]{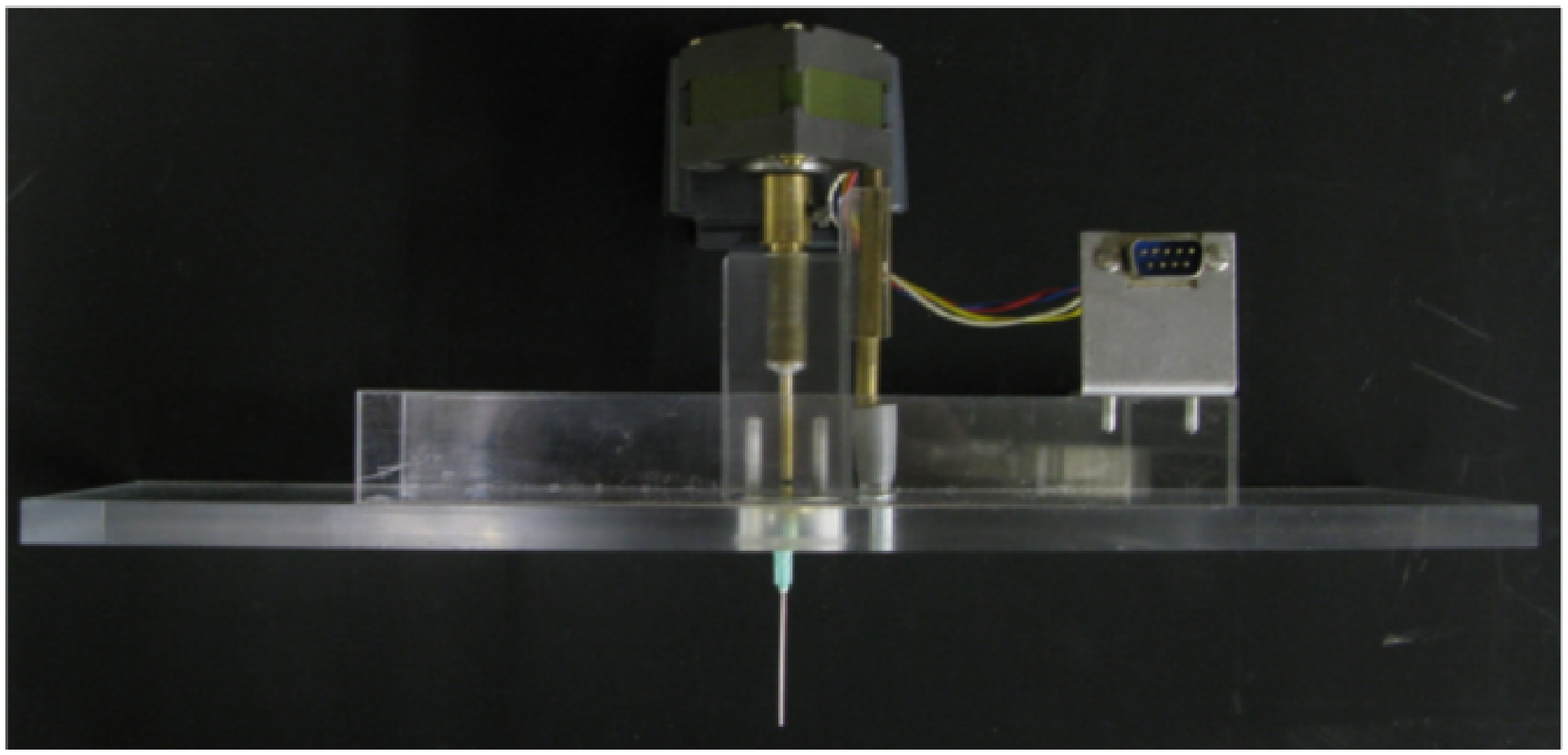}}
\end{center}
\caption{The experimental set-up: the drop generator with its driver, the container, and the devices for visualization (top). Photograph of the drop generator (bottom).}
\label{expset}
\end{figure}

Experiments were performed in a glass-wall vessel with the square cross-section $20$ cm x $20$ cm and the height $100$ cm (see Fig.~\ref{expset}).
The container was filled almost up to the top with $86\%$ glycerol of density $\rho$~=~1224~kg/m$^{3}$ and kinematic viscosity $\nu$~=~0.9$\cdotp$10$^{-4}$~m$^{2}$/s. The  corresponding dynamic viscosity $\eta \equiv \rho \nu = 110$ mPa s. 

A suspension was prepared by mixing glass particles with the $86\%$ glycerol (the fluid taken from the container). The volume fraction was set constant to $\phi=0.1$. To control the volume fraction, a certain quantity of glycerol was weighted to determine its mass $M$. The mass $M_{P}$ of glass particles required to obtain the volume fraction $\phi=0.1$ was calculated from the relation 
\bee
\phi = \frac{M_{P}/\rho_{P}}{M/\rho+M_P/\rho_{P}}, 
\eee
with 
the glass-particle density $\rho_{P} = 2400 \mbox{ kg/m}^{3}$. 

\begin{figure}[h]  
\begin{center}
{\includegraphics[width=8.6cm]{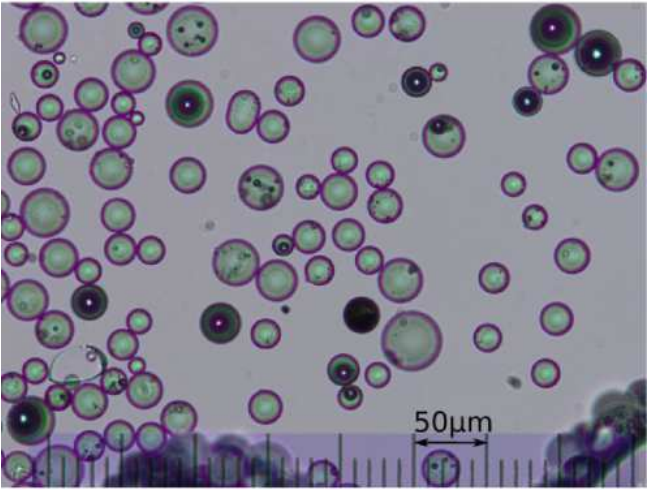}}
\end{center}
\caption{Glass particles under the optical microscope.}\label{czastki_mikroskop}
\end{figure}

The glass particles were polydisperse. The distribution of their radii was measured using a Nicon Eclipse $E50$ optical microscope equipped with a $20 \times$ magnifying lens.
A typical image observed under the microscope is shown in Fig.~\ref{czastki_mikroskop}. The actual size  of $411$ particles was measured, resulting in the mean particle diameter $d= (25 \pm 8 ) \mu \mbox{m}$.

A drop of the suspension was generated by a trigger mechanism, shown in the top panel of Fig.~\ref{expset}. This mechanism consisted of a pump-needle-unit (enlarged in the bottom panel of Fig.~\ref{expset}), a sequencer, a function generator and a power supply. The sequencer allowed to control the size of droplets. In the experiments, the drop diameter was close to $1$ mm. 

\begin{figure}[h]  
\begin{center}
{\scalebox{0.22}{\includegraphics{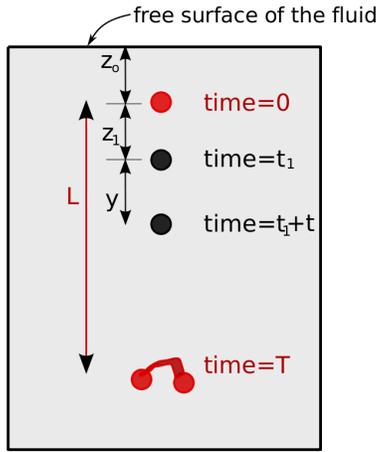}}}
\end{center}
\caption{Illustration of the experiments.}\label{rys}
\end{figure}

A single suspension drop was released at a distance $h$ from the side wall of the container and at the distance $z_{o}=4$ mm below the free surface of the fluid. Its motion and shape evolution were observed, as illustrated in Fig.~\ref{rys}, and recorded by a SONY XCD-X710 video camera (black and white). The position of the camera was changed stepwise to keep the drop within the field of view.

The goal of our investigations was to study the influence of a vertical wall on the motion, shape and break-up of a suspension drop. Therefore, we investigated the evolution of drops, which were released at different distances $h = 2, 5, 10, 20, 30, 50, 100$ mm between their centers and the wall. For a given value of $h$, the measurements were typically repeated $n=20-35$ times, to determine the dispersion of the results and reduce statistical errors. In total, $202$ measurements were performed. 

To control the initial size of all the drops, at each experiment the procedure of release was tested by measuring the time $t$ the drop needed to settle the same distance $y$, (with $y=30$ cm for $h=2$ mm and $h=5$ mm, and $y=50$ mm for all other distances $h$ from the wall), starting at $(z_{1}+z_{o})=41$ mm from the free surface of the fluid in the container, see Fig.~\ref{rys}. The actual initial diameter $D$ of each drop was later determined from photographs taken with the same SONY video camera during the experiments, with the average over all the measurements $\langle D \rangle = 1.13$ mm.

The Reynolds number based on $\langle D\rangle$ and the velocity $v=0.92$ mm/s of the drop in the central part of the container was much smaller than unity, $Re \!=\!\langle D \rangle v/\nu \!\approx \! 10^{-2}$.

The number of particles $N_0$ inside a single suspension drop of diameter $D$ was estimated in the following way.
First, the sizes of $n_{d}=411$ glass particles were measured under the microscope, and the total volume $V_{d}$ of these $n_{d}$ particles was calculated. 
Since the volume $V_D$ of all the $N_0$ particles inside a suspension drop of 
diameter $D$ and the volume fraction $\phi=0.1$ was known,
\bee
V_{D}&=&\phi \frac{\pi  D^{3}}{6},
\eee
then for each suspension drop, $N_0$ 
was determined from the rescaling relation,
\begin{equation}\label{nnn}
 N_{0}=\frac{V_{D} n_{d}}{V_{d}}.
\end{equation}

Finally, the number of particles in each suspension drop $N_0$ was averaged over all the experiments, with 
$\langle N_{0} \rangle = 7100 \pm 200$.

\section{Experimental results}\label{3}
\subsection{General observations}

During the experiments it was observed that a suspension drop settling close to a vertical wall evolved qualitatively in a similar way as  in an unbounded fluid. 
At the beginning, the drop remained almost spherical, leaving a thin tail of particles behind. The drop shape gradually changed into a horizontal torus, which continued to flatten and widen, and suddenly deformed and destabilized into two (or sometimes more) droplets. Therefore, the main characteristic features of the drop evolution observed previously in an unbounded fluid (see, e.g., Fig. 23 in Ref. [\onlinecite{art9}]
and Figs. 18 and 21 in Ref.~[\onlinecite{art21}]) 
 are also seen in the presence of a vertical wall. There are, however, the following differences. Close to a wall oriented parallel to the gravitational field, the drop looses more particles; also, it flattens and breaks up earlier. Moreover, the drop settling very close to the wall (at $h=2$ mm) destabilized in such a way that the line of centers of the two droplets formed after the break-up in most cases was oriented along the wall (in 23 out of 25 such experiments). At $h\ge 5$ mm, the lines of centers were oriented randomly.

The statistics of the observed evolution patterns is listed in Table~\ref{tata2}. All the drops broke up, 99\% of them into two fragments, and only two drops into three pieces.

\begin{table}[h]
\caption{The number of suspension drops with the indicated final stage of the evolution in our experiments.} \label{tata2}
\begin{center}
\begin{tabular}{|c|c|c|c|c|c|}\hline
{h/D} & {destabilization} & {destabilization} & {no break-up} & {total} \\ 
& into 2 droplets& into 3 droplets&&\\ \hline
$87.7$ & $34$ & $0$ & $0$ & $34$ \\ \hline
$43.1$ & $35$ & $1$ & $0$ & $36$ \\ \hline
$26.8$ & $33$ & $0$ & $0$ & $33$ \\ \hline
$17.9$ & $27$ & $1$ & $0$ & $28$ \\ \hline
$8.6$ & $22$ & $0$ & $0$ & $22$ \\ \hline
$4.6$ & $21$ & $0$ & $0$ & $21$ \\ \hline
$1.8$ & $24$ & $0$ & $0$ & $24$ \\ \hline
\end{tabular}
\end{center}
\end{table}

\subsection{Dependence of destabilization time and length on the distance from the wall}\label{expde}
In the experiments, we measured the instant $T$ of drop destabilization, and the vertical drop position $L$ at that time, as functions of the distance $h$ from the wall. 
The destabilization time $T$ was defined 
as the moment when the torus began to break, and the destabilization length $L$ as the distance traveled by the drop from its release (corresponding to time equal to zero) until destabilization, see Fig.~\ref{rys}. Breaking of the torus was observed on the side view. Just before destabilization, the torus became thinner in the central part than at the outer ones. Then, it began to bend, with the thinner central part above the thicker outer ones, and this moment was called the destabilization time. 
For each drop, $T$ and $L$ were measured using a stopwatch and a scale on the back wall of the container. Then, values of $T$ and $L$ were averaged over all $n$ experiments performed for a given distance $h$. 

Values of the average destabilization time and length for different distances $h$ are listed in Table~\ref{T1}. For simplicity, we use here the same symbols $T$ and $L$, which from now on in this paper will denote the average values.
\begin{table}[h]
\caption{The destabilization time $T$ and length $L$ of suspension drops as functions of the distance from the wall $h$, averaged over all experiments performed for a given distance $h$. The average initial drop diameter $D$ and the inverse time unit, $1/\tau$, defined in Eq.~\eqref{tau}, are also indicated.} \label{T1}
\begin{center}
\begin{tabular}{|c|c|c|c|c|}\hline
$h$ [mm]& ${D}$ [mm] & ${L}$ [mm]& ${T}$ [s]& $1/\tau $ [1/s]\\ \hline 
$2$  & $1.14$ & $186 \pm 10$ & $348 \pm 19$ & $1.59$ \\ \hline
$5$  & $1.08$ & $266 \pm 9$  & $434 \pm 16$ & $1.51$ \\ \hline
$10$ & $1.16$ & $326 \pm 8$  & $501 \pm 10$ & $1.62$ \\ \hline
$20$ & $1.12$ & $319 \pm 10$ & $477 \pm 13$ & $1.56$ \\ \hline
$30$ & $1.12$ & $333 \pm 7$  & $467 \pm 11$ & $1.57$ \\ \hline
$50$ & $1.16$ & $344 \pm 8$  & $466 \pm 9$  & $1.61$ \\ \hline
$100$ & $1.14$ & $342 \pm 8$ & $488 \pm 10$ & $1.59$ \\ \hline
\end{tabular}
\end{center}
\end{table}

The actual initial diameter $D$ of each drop was determined from photographs taken during the experiment at a time instant between 3 s and 12 s, and averaged over the experiments performed at a given distance $h$. If the diameter of a drop was more than 15 percent different from the average, such an experiment was discarded. Then, the average diameter was recalculated, excluding the discarded experiments, with the standard error of the mean equal to 0.01mm. The resulting values $D$ vary a little with a change of $h$, because of statistical fluctuations. 

To compare the experimental data, we therefore used dimensionless variables with $h$-dependent units - 
the drop initial diameter $D$, its initial settling velocity 
\begin{equation}
 \upsilon = \frac{2F_{tot}}{5 \pi \eta D},\label{uu}
\end{equation}
\\
determined by the total gravitational force (weight minus buoyancy) acting on the drop,
\begin{equation}
 F_{tot}=\frac{\pi D^{3} \phi}{6} (\rho_{P}-\rho) g,
\end{equation}
and the corresponding time unit,
\bee
 \tau &=& \frac{D}{2\upsilon}\label{tau},
\eee
or explicitly, $\tau = 15 \eta [2 g (\rho_{P}-\rho) D \phi]^{-1}$.
Here $g$ is the gravitational acceleration.

\begin{figure}[ht]  
\begin{center}
\includegraphics[width=9.4cm]{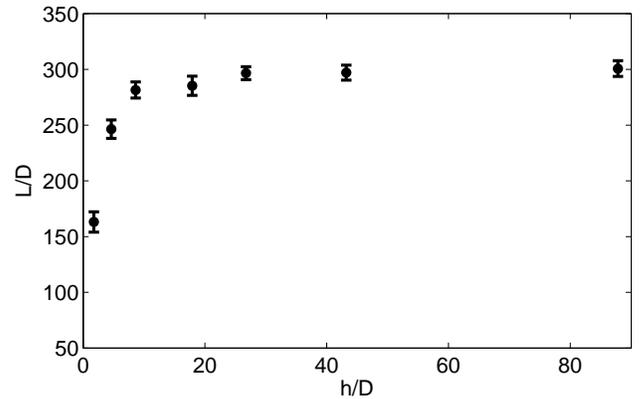}
\end{center}
\caption{Experimental results: The average drop destabilization length versus the distance from the wall, both normalized by the initial diameter of the drop.} \label{Lh}
\end{figure}

\begin{figure}[ht]  
\begin{center}
{\includegraphics[width=9.4cm]{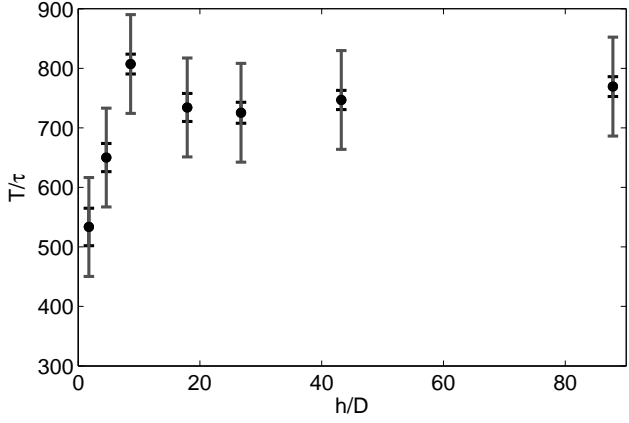}}
\end{center}
\caption{Experimental results: The average drop destabilization time $T/\tau$ versus the distance from the wall $h/D$. Statistical errors (black) and systematic errors (gray) are indicated.} \label{7}
\end{figure}

In Figs. \ref{Lh} and \ref{7}, the average non-dimensional destabilization length $L/D$ and time $T/\tau$ are plotted versus the non-dimensional distance $h/D$ of the drop center from the wall. In both figures, the statistical errors are indicated by the black error bars. The measurements of time $T/\tau$ have an additional ``systematic`` error, related to uncertainty of $\tau$, owing to fluctuations of the initial drop velocity $\upsilon$ from day to day. They might be related to small temperature variations of the fluid, temperature gradients, convection of the fluid, water at the interface of the glycerol, or other reasons. 
The method to estimate such a systematic error, shown in Fig.~\ref{7} by the gray error bars, is described in Appendix \ref{error}.

It is clear that the destabilization time and length of a drop are shorter for smaller distance $h$ from the wall. In particular, the differences between values attained at $h=2$ mm and $h=10$ mm are significantly larger than both statistical and systematic errors. 
In Sec.~\ref{compare}, the results of the measurements will be extensively discussed and compared with results from computations based on the point-particle model.

\section{Modeling the internal structure of suspension drops}\label{3.5}
In the experiments, the suspension of particles is polydisperse, as shown in Fig.~\ref{czastki_mikroskop}, and the average number $N_0$ of particles in a drop is 
greater than 7000. 
Numerical computation of the particle dynamics during a very long time, taking into account the size distribution, and for a large number of initial random configurations of particles, located at different distances from the wall, would be very time consuming and complicated. Our goal is to concentrate on essential features of the drop evolution, therefore we simplify the system. 
At the beginning of the experiments, a suspension drop is spherical with diameter $D$, it consists of $N_0$ particles of diameters $d_i$, with $i=1,...,N_0$, and an average of 25$\mu$m, and the drop is subject to the total gravitational force (weight minus buoyancy) $F_{tot}$.

In the numerical simulations, we use a simplified model of the drop internal structure. The average initial diameters of a model drop and the real one are the same, $D=1.13$mm, and the total gravitational forces are equal, $F_{tot}$. However, a model drop consists of identical particles of diameter $d_m$. 
Moreover, to simplify the computations, the initial number of particles inside the model drop, $N$, is one order of magnitude smaller than inside the real one, $N_0$.

To compensate for this difference, there are different possible choices of the particle diameter $d_m$. In model I, we require that the volume fraction $\phi_m$ of the model drop is the same as the volume fraction $\phi_0$ of the real drops,
\bee
\mbox{Model I: }&& \;\phi_m = \phi_0,
\eee
or equivalently,
\bee
N\left( \frac{d_m}{D}\right)^3= 0.1. \label{mI} 
\eee

In model II, we assume equal mean diameters of particles inside the model and the real drop, 
\bee
\mbox{Model II:}&& \;d_m = \frac{1}{N_0} \sum_{i=1}^{N_0} d_i.\label{mII}
\eee

\begin{figure}[ht]  
\begin{center}
{\includegraphics[width=6cm]{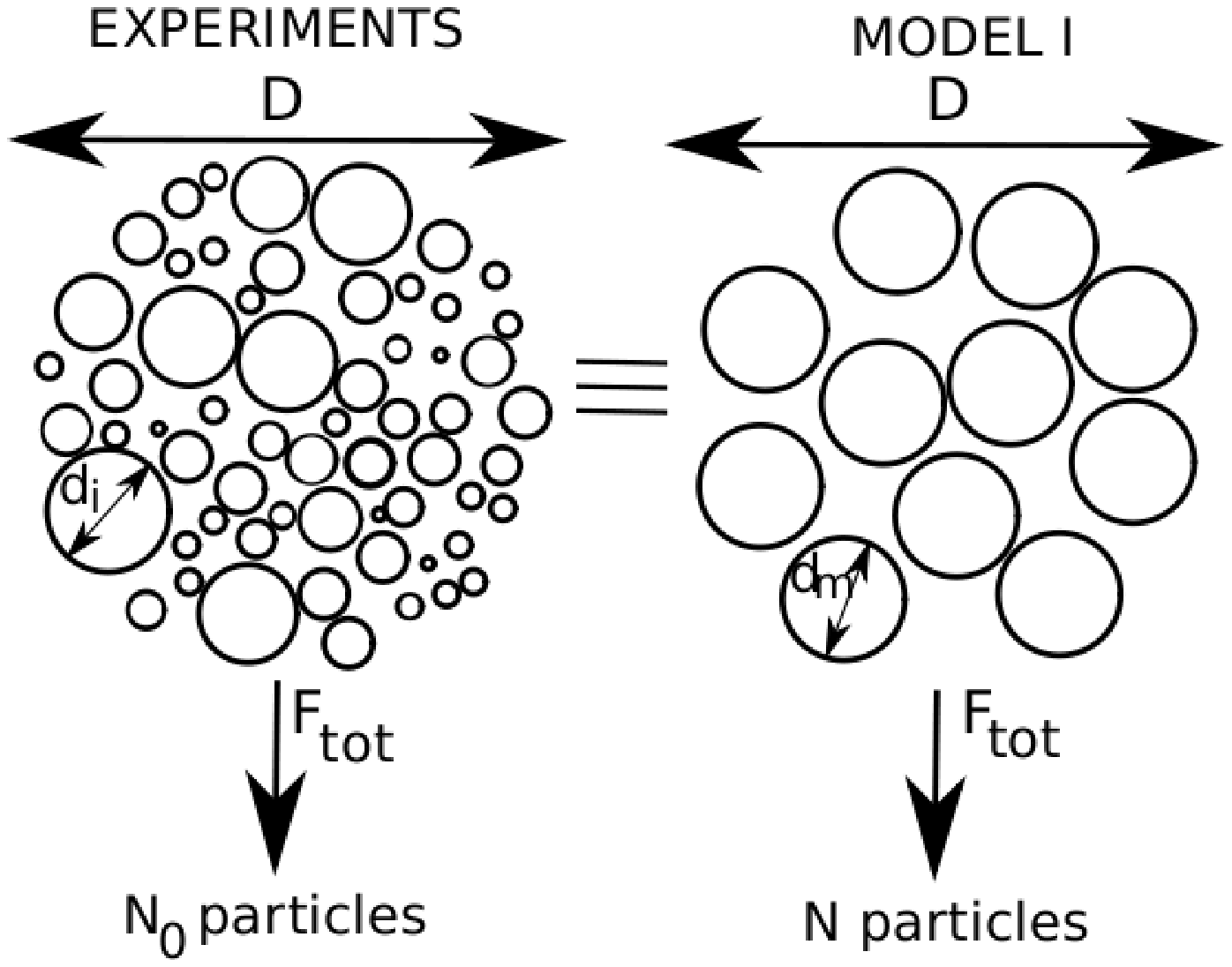}}
\caption{Model I of the drop internal structure: equal volume fractions.} \label{ism1}

\vspace{0.7cm}
{\includegraphics[width=6cm]{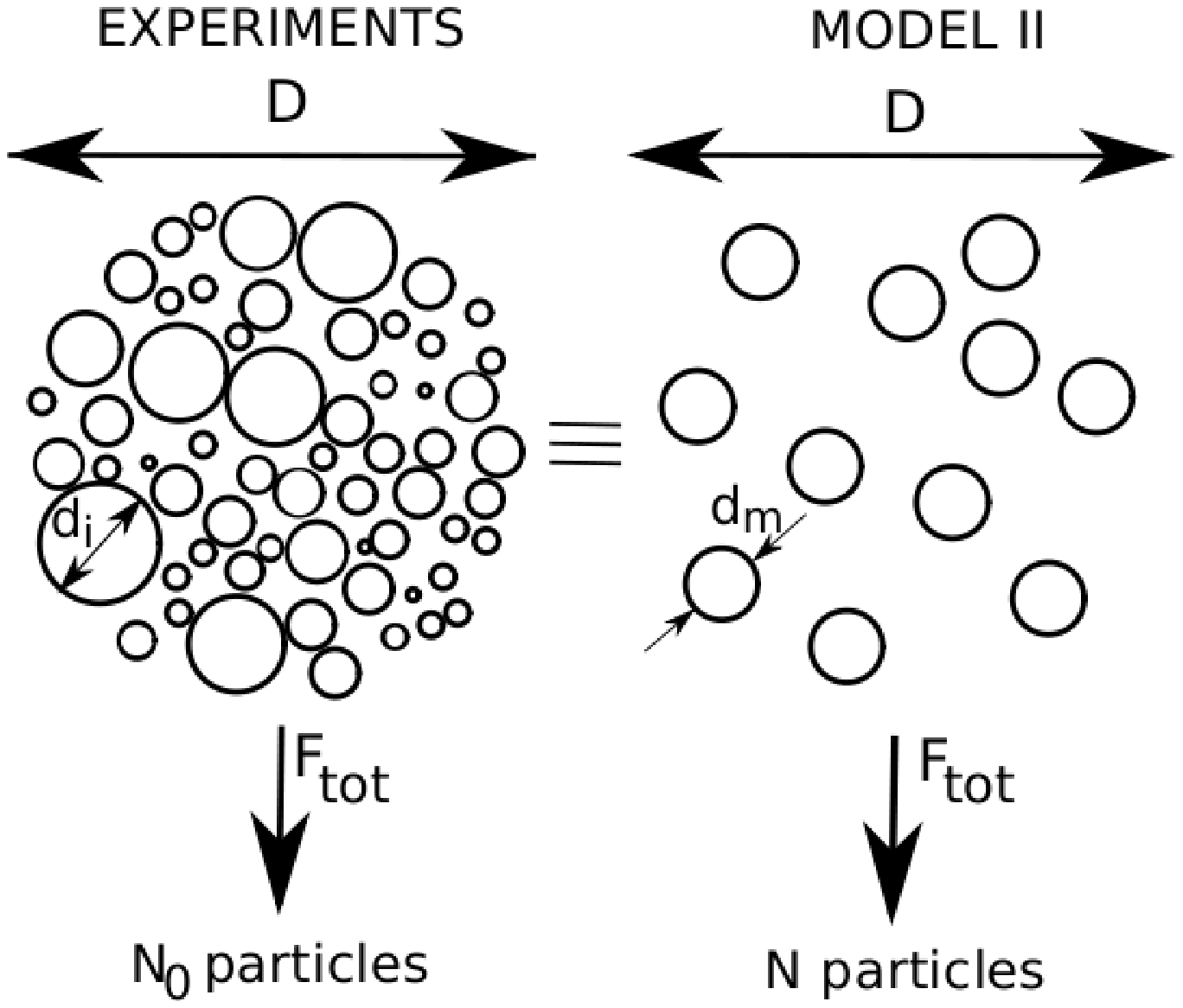}} 
\end{center}
\vspace{-0.15cm}
\caption{Model II of the drop internal structure: equal mean particle diameters.} \label{ism2}
\end{figure}

Both models are schematically shown in Figs.~\ref{ism1} and \ref{ism2}. 

Other values of $d_m$, between those following from the model I and II, are also reasonable. 
The point is that the dynamics of the model drop only weakly depends on the choice of the particle diameter $d_m$: as in model I, II or in between. To illustrate this property, we calculate 
the Stokes velocity of a single particle in a model drop, 
\bee
{V_0} &=& \frac{F_{tot}}{3\pi \eta d_m}.
\eee
Then, we compare it to the initial drop velocity $\upsilon$, given by Eq.~\eqref{uu}.
With the use of Eqs.~\eqref{mI} and \eqref{mII}, 
\bee
\frac{V_0}{\upsilon} = \frac{5}{6Nd_m/D} =\left\{
\begin{array}{l l}
0.02 &  \mbox{for model I,}\\
0.05 &  \mbox{for model II}.
\end{array} \right.
\eee
Therefore, in both models, and also for intermediate values of $d_m$, 
the particle Stokes velocity $V_0$ is very small in comparison to 
the initial settling velocity of the drop $\upsilon$. A specific choice of $d_m$ is not relevant.

\section{Theoretical results}\label{4}
In this paper, 
the theoretical approach is based on the point-particle 
model. In this way, the evaluation of the particle dynamics is much simpler numerically and has a smaller number of parameters than, e.g., in case of the accurate multipole algorithm for spherical particles.\cite{Cichocki-Ekiel_Jezewska-Wajnryb:1999} The main advantage is that the point-particle dynamics does not depend on the particle radius, if it is carried out in the frame of reference moving with the Stokes velocity~$V_0$.

\subsection{Point-particle model close to a plane solid wall}

In a system with Reynolds number much smaller than unity (as in our experiments), the fluid velocity ${\bf u}({\bf r})$ and pressure $p({\bf r})$ can be determined from the Stokes equations. The fluid flow generated by motion of particles, which are subject to external non-hydrodynamic forces (e.g. gravitational ones), is often described within the point-particle model.\cite{Kim-Karrila:1991,Happel-Brenner:1986} 
In this case, the Stokes equations take the form
\bee
\eta \nabla^2 {\bf u}({\bf r}) - {\bm \nabla} p({\bf r}) &=& - \sum_{\alpha=1}^N {\bf F}_{\alpha} \delta({\bf r}-{\bf r}_{\alpha}),\\
{\bm \nabla} \cdot {\bf u}({\bf r})&=&0\eee
where ${\bf r}_{\alpha}$ and ${\bf F}_{\alpha}$ are the position of a particle $\alpha$, and the external force it exerts on the fluid, respectively. Here $N$ is the number of particles, and $\eta$ is the dynamic viscosity of the fluid. The solution is the sum of the Green tensors appropriate for the specific geometry of the system and the boundary conditions. For an unbounded fluid, these conditions state that $|{\bf u}({\bf r})|\rightarrow 0$ when $|{\bf r}|\rightarrow \infty$.
In addition, for a fluid bounded by a solid wall located at $z=0$, the stick-boundary condition has to be satisfied at the wall, i.e. $|{\bf u}({\bf r})|=0$ if ${\bf r}=(x,y,0)$.

\begin{figure}[ht]  
{\includegraphics[width=8.6cm]{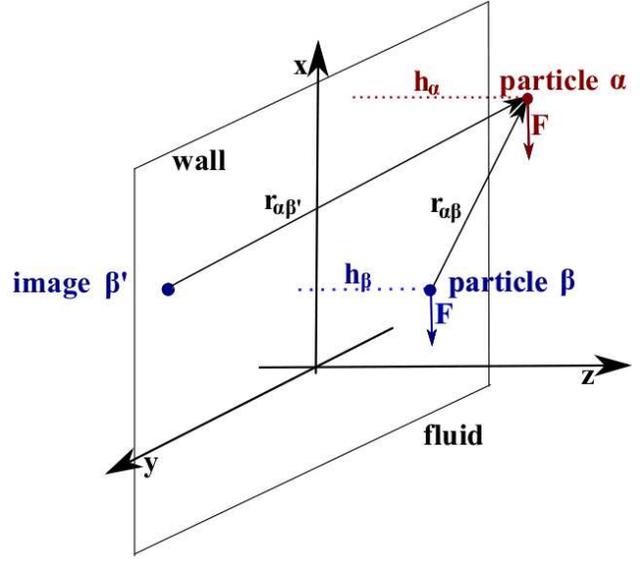}}
\caption{The system, the coordinates and the notation.}\label{notation}
\end{figure}

We assume that all the point forces are identical, ${\bf F}_{\alpha}$~=~${\bf F}$, and parallel to the solid wall. We choose a system of coordinates in which ${\bf F}=(-F,0,0)$, with $F>0$, 
as illustrated in Fig.~\ref{notation}.

We choose the frame of reference moving with the Stokes velocity of a single point-particle in an unbounded fluid. In this way, the dynamics of particles and the fluid flow are independent of the particle radius.
When the wall is present, the point-particle motion is determined by the following expressions for the particle velocities, $\alpha=1,...,N$,
\begin{equation}
\textbf{v}_{\alpha} = \sum_{\beta \neq \alpha}^{N} \textbf{T}(\textbf{r}_{\alpha},\textbf{r}_{\beta}) \cdot \textbf{F} + \tilde{\textbf{T}}(\textbf{r}_{\alpha\alpha'}) \cdot \textbf{F}.\label{vwall}
\end{equation}
Here the Blake tensor \cite{Blake:1971}
\bee
\textbf{T}(\textbf{r}_{\alpha},\textbf{r}_{\beta})  &=& \textbf{T}_{0}(\textbf{r}_{\alpha \beta})  + \tilde{\textbf{T}}(\textbf{r}_{\alpha\beta'})
\eee
is the sum of the Oseen tensor (the Green tensor in the unbounded fluid),
\bee
\textbf{T}_{0}(\textbf{r}
_{\alpha \beta}
) &=& \frac{1}{8\pi \eta r_{\alpha\beta}}\left({\bf I} +\frac{{\bf r}_{\alpha\beta}{\bf r}_{\alpha\beta}}{r_{\alpha\beta}^2} \right),
\eee
and the tensor $\tilde{\textbf{T}}(\textbf{r}_{\alpha\beta'})$, defined as
\bee
\tilde{\textbf{T}}(\textbf{r}_{\alpha\beta'}) \cdot \textbf{F} &=&  - \textbf{T}_{0}(\textbf{r}_{\alpha \beta'}) \cdot \textbf{F}\\
&&-2h_{\beta} \textbf{F} \cdot \textbf{P} \cdot {\bm \nabla}_{{\bf r}_{\alpha \beta'}} \textbf{T}_{0}(\textbf{r}_{\alpha \beta'}) \cdot \hat{\bf z}
\\
&+&\dfrac{2h_{\beta}^2}{8 \pi \eta} \textbf{F} \cdot \textbf{P} \cdot {\bm \nabla}_{{\bf r}_{\alpha \beta'}} \left( \dfrac{ \textbf{r}_{\alpha \beta'}}{r_{\alpha \beta'}^3} \right).
\eee
The Oseen tensor, $\textbf{T}_{0}(\textbf{r}
_{\alpha \beta}
)$, describes the interaction of a particle $\alpha$ with the particle $\beta$ in an unbounded fluid, and the tensor $\tilde{\textbf{T}}(\textbf{r}_{\alpha\beta'})$ describes the interaction of a particle $\alpha$ with the mirror image $\beta'$ of the particle $\beta$.
Here $ \textbf{P}= \textbf{1}-2\hat{\bf z}\hat{\bf z}$ is the reflection operator, $h_{\beta}$ is the distance between the particle $\beta$ and the wall, ${\bf I}$ the unit tensor, $\hat{\bf z}$ the unit vector perpendicular to the wall, $\textbf{r}_{\alpha\gamma}=\textbf{r}_{\alpha}-\textbf{r}_{\gamma}$, and $r=|\textbf{r}|$. The tensor $\tilde{\textbf{T}}(\textbf{r})$  accounts for the difference between the Green tensor for the fluid bounded by the wall and unbounded.

In Eq.~\eqref{vwall}, which defines the velocity of a point-particle $\alpha$ close to a wall, the first term is the sum of the fluid velocity fields generated by all the other particles at the position $\textbf{r}_{\alpha}$ where the particle $\alpha$ is located. 
This term describes advection of a particle by the fluid flow generated by all the other particles. The second term is the self contribution: it specifies the velocity of a single particle interacting with the wall, in the frame of reference moving with the Stokes velocity of a single point-particle in an unbounded fluid. Since the wall is parallel to the force $\textbf{F}$, it follows that
\bee
\tilde{\textbf{T}}(\textbf{r}_{\alpha\alpha'}) \cdot \textbf{F} &=& -\frac{3 \textbf{F}}{32h_{\alpha} \pi \eta}.\label{imagealpha}
\eee

The dynamics of the point-particles $\alpha=1,...,N$ is governed by the following system of first order ODEs
\begin{equation}
\frac{d\textbf{r}_{\alpha}}{dt} = \textbf{v}_{\alpha},
\end{equation}
with the dependence of $\textbf{v}_{\alpha}$ on the positions $\textbf{r}_{\beta}$ of all particles $\beta=1,...,N$, given by Eq.~\eqref{vwall}.

\subsection{Evolution of a suspension drop}
In this section, we apply the point-particle model to describe the evolution of a suspension drop, with the same fluid outside and inside. $N$ identical point forces ${\bf F}$ are distributed at random with a uniform $N$-particle probability distribution inside a spherical volume of diameter D. In this paper, $N=700$. Thirty such initial random configurations have been generated. 

The particle dynamics was determined based on the point-particle model close to a vertical wall, described in the previous section. The model was implemented numerically in our MATLAB code, using the variable order Adams-Bashforth-Moulton solver ode113 with the relative accuracy of 0.1\% and the absolute error of $10^{-6}$.
In the numerical simulations, the same set of initial random configurations was evolved for 
ten different distances $h/D$ of the drop center from the wall (see Table~\ref{tab2} for the list of values), and for a drop in an unbounded fluid. 

We introduce dimensionless quantities by taking $D$ as the length unit, and $\upsilon$, given in Eq.~\eqref{uu}, with
\bee
F_{tot}=NF,\label{fd}
\eee
as the velocity unit, with $F=|{\bf F}|$. The corresponding time unit $\tau$ is defined by Eqs.~\eqref{tau} and \eqref{fd}. It is known from the literature \cite{art10,FF} that, in an unbounded fluid of viscosity $\eta$, the settling velocity of a suspension drop of diameter $D$, subject to $N$ identical, randomly distributed point-forces $F$, is well approximated by the expressions~\eqref{uu} and \eqref{fd}, if $N$ is sufficiently large. For arbitrary $N$, the exact theoretical expressions were obtained in Ref.~[\onlinecite{art26}] 
 by statistical averaging. An isolated suspension drop settles with the same velocity $\upsilon$ as a fluid drop of the same size and the same excess weight. 

In the simulation, we evaluated the dynamics of the particles in each suspension drop until $T_{k}=3920 \tau$.
The generic pattern of the drop evolution was the same as in an unbounded fluid \cite{art21,art9} and in our experiments. The statistics  is shown in Table \ref{tab2}. 
Initially, the suspension drops were spherical, with the particles randomly distributed inside their volume. Later, the drops gradually lost single particles, which were left behind in a thin tail. During the motion, the drops expanded horizontally and contracted vertically, in a form of a horizontal torus. All the time, the particles recirculated inside the drop, moving faster in the inner parts of the drop, slower in the outer parts, coming back to the inner parts again, and so on. Suddenly, the torus bended, and broke into two (or sometimes three) fragments, which became spherical and repeated the generic evolution pattern. Such a scenario happened for 77\% of the total 330 suspension drops simulated numerically. In 98\% of the destabilization events, the drop broke into two fragments; only 5 drops out of 254 broke into three pieces. 23\% of the drops did not break up before $T_{k}=3920 \tau$. Some of the particles were gradually lost from these drops, one by one, while the other particles inside the drops recirculated all the time. In Ref.~[\onlinecite{art9}], 
 it was shown that in unbounded fluid, the percentage of particle clouds which break up increases with the increase of the initial number $N$ of point-particles inside the cloud. 
\begin{table}[h]
\caption{The number of suspension drops with the indicated final stage of the evolution at $T_{k}\!\!=\!\!3920\tau$ (the point-particle model).} \label{tab2}
\begin{tabular}{|c|c|c|c|c|c|}\hline
{h/D} & {destabilization} & {destabilization} & {no break-up} & {total} \\ 
& into 2 droplets& into 3 droplets&&\\ \hline
$\infty$ & $24$ & $0$ & $6$ & $30$ \\ \hline
$70$ & $16$ & $1$ & $13$ & $30$ \\ \hline
$30$ & $22$ & $1$ & $7$ & $30$ \\ \hline
$10$ & $24$ & $0$ & $6$ & $30$ \\ \hline
$6.5$ & $24$ & $0$ & $6$ & $30$ \\ \hline
$5$ & $22$ & $0$ & $8$ & $30$ \\ \hline
$3.5$ & $24$ & $1$ & $5$ & $30$ \\ \hline
$2.5$ & $26$ & $1$ & $3$ & $30$ \\ \hline
$1.5$ & $19$ & $1$ & $10$ & $30$ \\ \hline
$1$ & $22$ & $0$ & $8$ & $30$ \\ \hline
$0.75$ & $26$ & $0$ & $4$ & $30$ \\ \hline
\end{tabular}
\label{sre1}
\end{table}

In the following, we will discuss only the evolution of the drops which broke up. Analyzing the results of the numerical simulations, we use the same definition of 
the destabilization time $T$ as in the experiments, see Sec.~\ref{expde}. Within the point-particle model, the destabilization length $L-V_0T$ corresponds to the distance the drop traveled during time $T$, in the frame of reference moving with the Stokes velocity $V_0$ of a single particle in an unbounded fluid. 

\begin{figure}[h]  
\begin{center}
{\includegraphics[width=9.1cm]{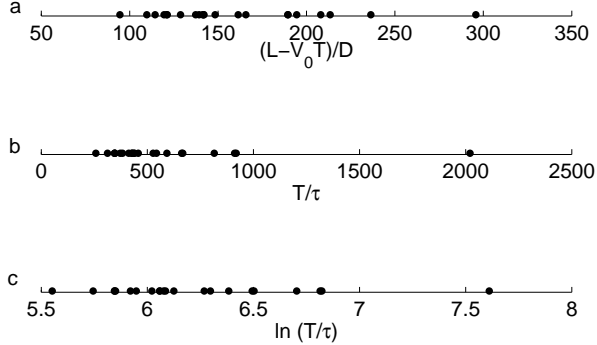}}
\end{center}
\vspace{-0.9cm}
\caption{
Destabilization lengths (a) and times (b,c) of individual drops, for $h/D=5$.
} \label{statystyka}
\end{figure}

A typical statistics of destabilization lengths and times is shown  in 
Fig.~\ref{statystyka}, where the initial distance of the drop center from the wall $h/D=5$.
From Fig.~\ref{statystyka}a and b it is clear that destabilization lengths, $(L-V_0T)/D$, and times, $T/\tau$, of individual drops differ significantly from one suspension drop to the other. $T/\tau$ can even change as much as almost one order of magnitude. Fluctuations of $(L-V_0T)/D$ are smaller, but still the values can differ even by a factor of 3. This effect is related to the chaotic nature of many-particle dynamics.\cite{art1} 

Another striking feature is the non-symmetric distribution of destabilization times, which more frequently attain smaller values, see Fig.~\ref{statystyka}b. This asymmetry is reduced if we look at the statistics of the natural logarithm $\ln(T/\tau)$ rather than $T/\tau$, as shown in Fig.~\ref{statystyka}c. Therefore, the average destabilization time is evaluated as the exponential function of the mean $\ln(T/\tau)$ throughout this paper (both in the numerical simulations and the experiments).

\begin{figure}[h]  
{\includegraphics[width=9.1cm]{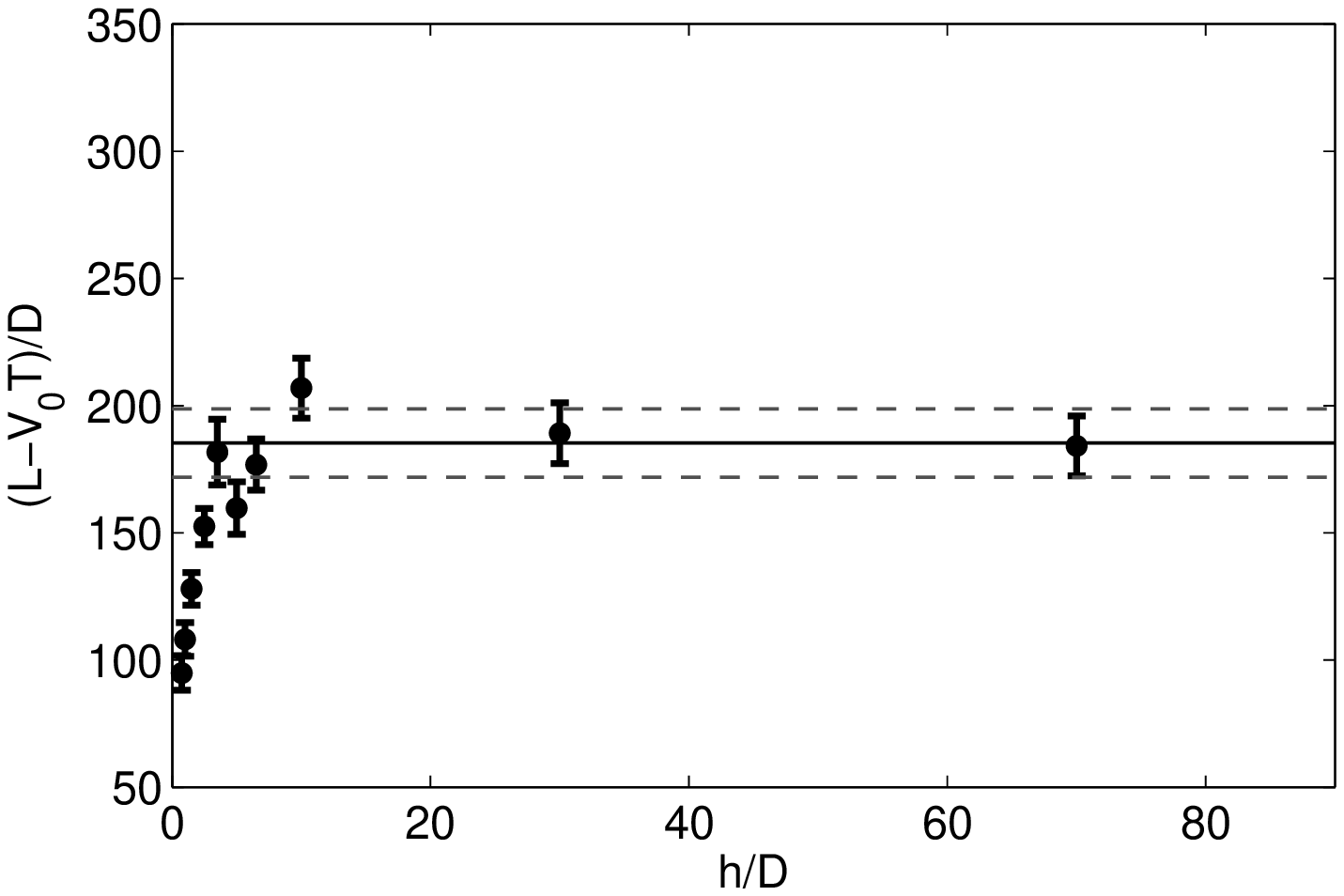}}
\vspace{-0.5cm}
\caption{The average drop destabilization length $(L-V_0T)/D$ versus the distance from the wall $h/D$. The horizontal solid line and dashed lines correspond to the average destabilization length and the standard error of the mean, for an unbounded fluid.} \label{sLh}
%
{\includegraphics[width=9.1cm]{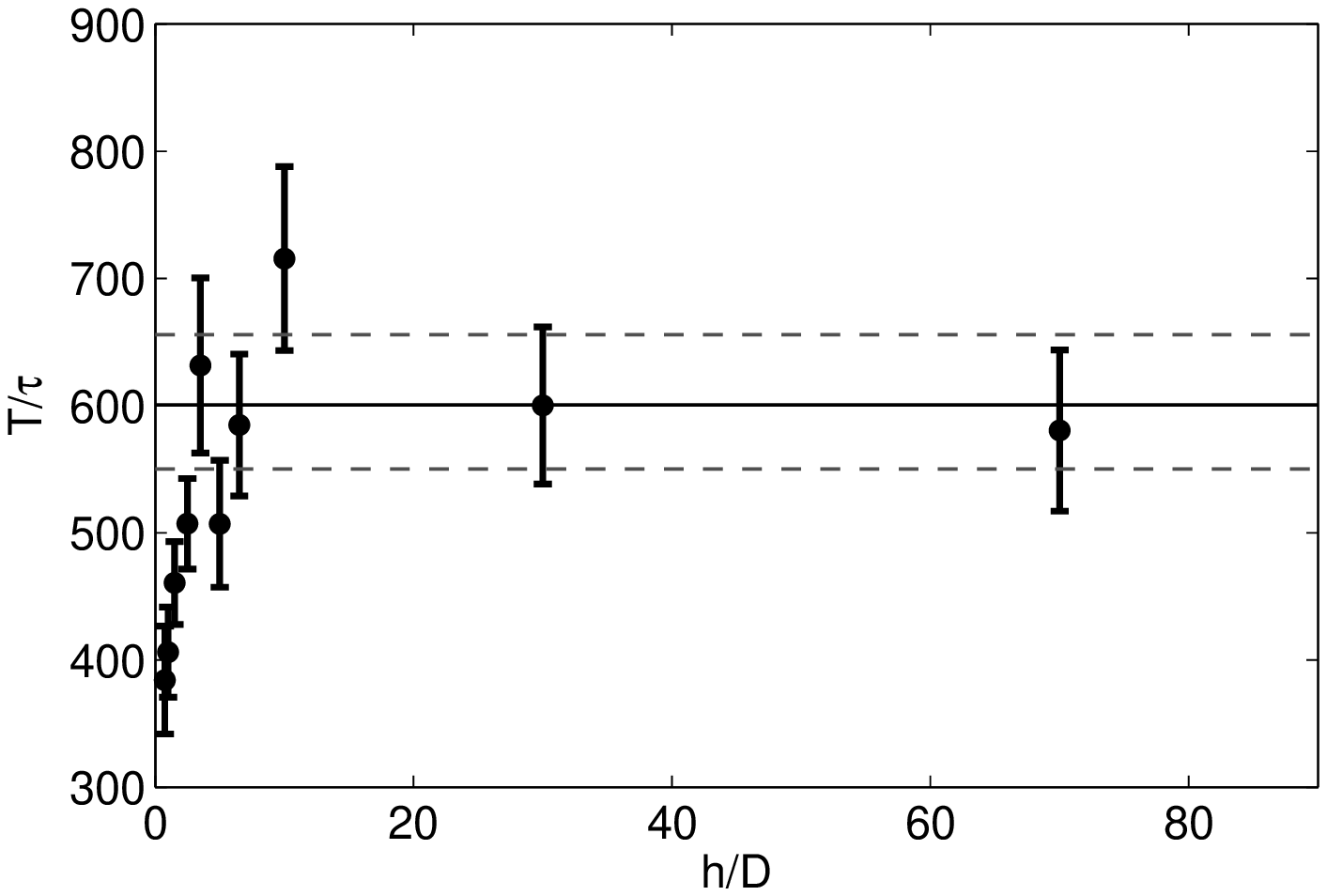}}  
\vspace{-0.5cm}
\caption{The average drop destabilization time $T/\tau$ versus the distance from the wall $h/D$. The horizontal solid line and dashed lines correspond to the average value of $T/\tau$ and standard error of the mean, for an unbounded fluid.} \label{s7}
\end{figure}

The average numerical values of the destabilization length and time are listed in Table~\ref{tabIV},  
and plotted in Figs.~\ref{sLh}-\ref{s7}. Again, for simplicity, we use the same symbols $L$ and $T$ there to denote the mean values. 
\begin{table}[h]
\caption{The average destabilization length, $(L-V_0T)/D$, and time, $T/\tau$, following from the point-particle model, for different distances from the drop center to the wall, $h/D$.} \label{tabIV}
\begin{tabular}{|c|c|c|c|}\hline
$h/D$ & $(L-V_0T)/D$ & $T/\tau$ \\ 
\hline
\hline

$\infty$ & $185 \pm 27$  & $601 \pm  53$  \\ \hline
$70$ & $184 \pm 24$ &  $580 \pm 63$ \\ \hline
$30$ & $189 \pm 24$ &  $600 \pm 62$ \\ \hline
$10$ & $207 \pm 24$ & $716 \pm 72$ \\ \hline
$6.5$ & $177 \pm 20$ & $585 \pm 56$ \\ \hline
$5$ & $160 \pm 21$ & $507 \pm 50$ \\ \hline
$3.5$ & $182 \pm 26$ & $632 \pm 69$ \\ \hline
$2.5$ & $153 \pm 14$ & $507 \pm 36$ \\ \hline
$1.5$ & $128 \pm 13$ & $461 \pm 33$ \\ \hline
$1$ & $108 \pm 13$& $406 \pm 36$ \\ \hline
$0.75$ & $95 \pm 14$ & $384 \pm 42$ \\ \hline
\end{tabular}
\end{table}
The averages are taken over values corresponding to suspension drops which destabilized during the numerical computation, i.e. at $T<T_k=3920\tau$, separately for each distance $h$ from the wall. Both destabilization length and time are smaller if the drop is closer to the wall. 

%

\section{Comparison of the experiments with the point-particle model}\label{compare}
\begin{figure*}[t]  
{\includegraphics[width=18.1cm]{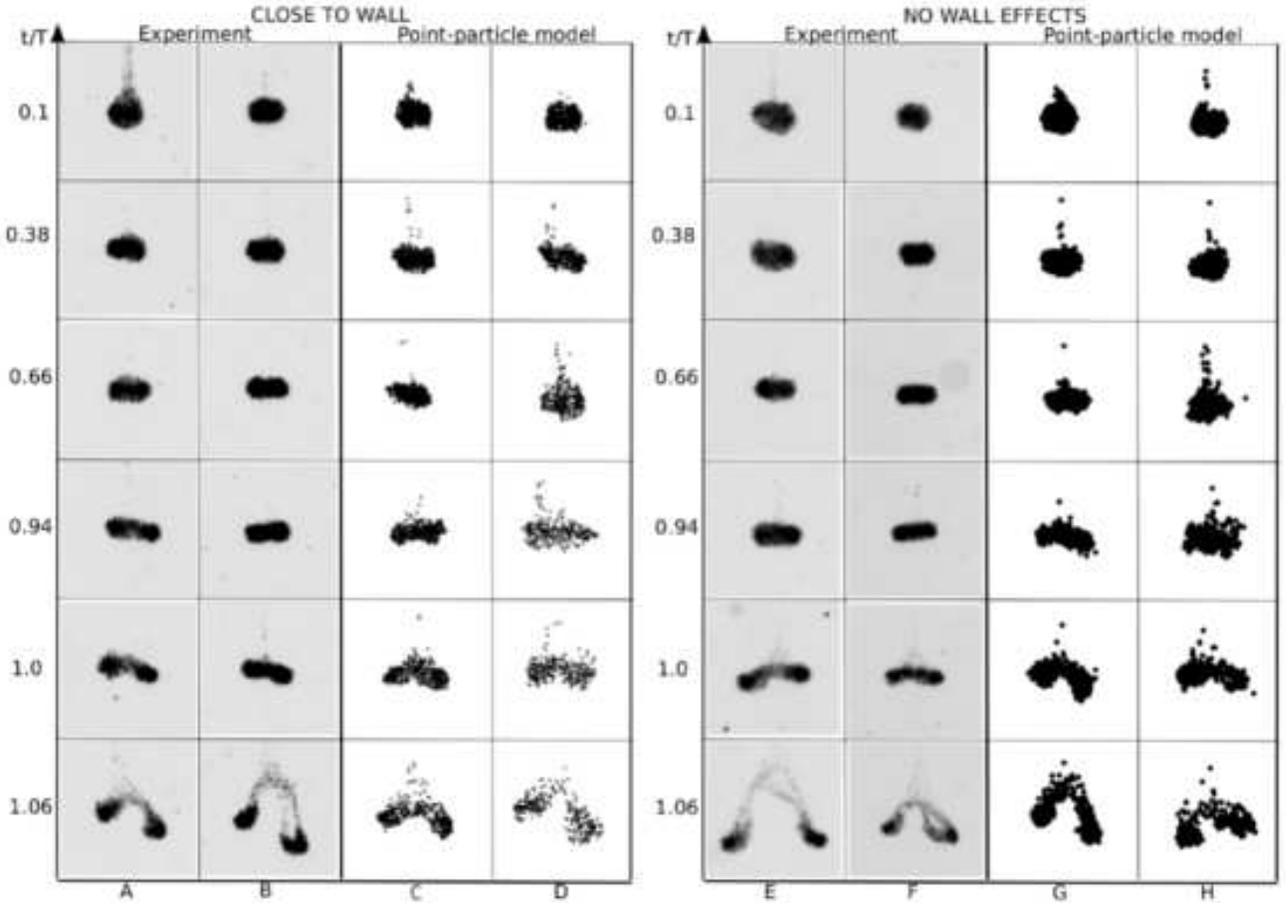}}
\caption{Shape evolution of suspension drops settling in a fluid close to a vertical wall (left, A-D), and very far from a wall (right, E-H). A and B: Experiments at $h/D=1.75$, with destabilization times $T/\tau=220$ and $T/\tau=634$,  respectively. C and D: Point-particle simulations at $h/D=1.5$, with 
$T/\tau=321$ and $T/\tau=966$, respectively. E and F: Experiments at $h/D\approx 88$, with destabilization times $T/\tau=478$ and $T/\tau=667$, respectively. G and H: Point-particle simulations in unbounded fluid, with 
$T/\tau=274$ and $T/\tau=1211$, respectively.}
\label{rys_dla_nn}
\end{figure*}

A typical evolution of 8 suspension drops is shown in Fig.~\ref{rys_dla_nn}. 
For each drop, time is normalized by the 
destabilization time of this specific drop. In these units, the time separations between the subsequent frames 1-4 are equal to 0.28, and between the subsequent frames 4-6 are smaller, equal to 0.06, to show more precisely the drop just before and after the break-up. All the frames have the same size 4D x 4D. The gravitational field is vertical.

There are 4 pairs of drops shown in subsequent columns, A-B, C-D, E-F, and G-H. In each pair, the drops differ only by the initial configuration of the particles. The two drops in the pairs
were selected such that the first one
breaks up at a significantly shorter time $T/\tau$ than the second one. The evolution patterns of the drops from the same pair are practically the same,
if observed in time $t$ normalized by the destabilization time $T$.

In Fig.~\ref{rys_dla_nn}, we compare 4 experiments (highlighted in gray) with the corresponding 4 numerical simulations based on the point-particle model (white). The shape evolution observed experimentally is well approximated by the corresponding shape evaluated from the point-particle model. The shape evolution is quite similar for drops close to a vertical wall and far from it. In the experiments, the drop just before break-up is a bit more flat and wide when the distance from the wall is larger. 

\begin{figure*}  
\begin{center}
{\includegraphics[width=18cm]{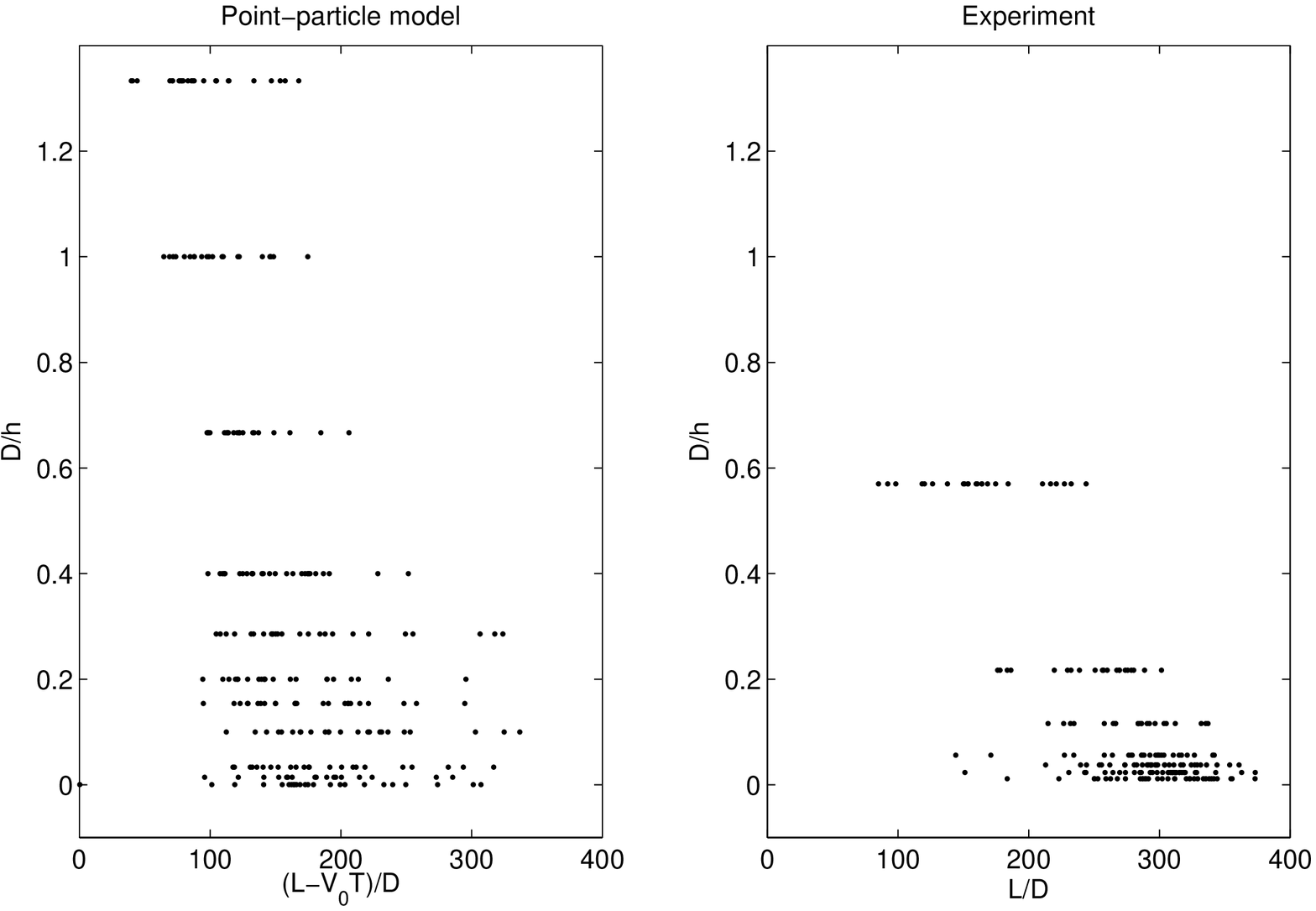}}
{\includegraphics[width=18cm]{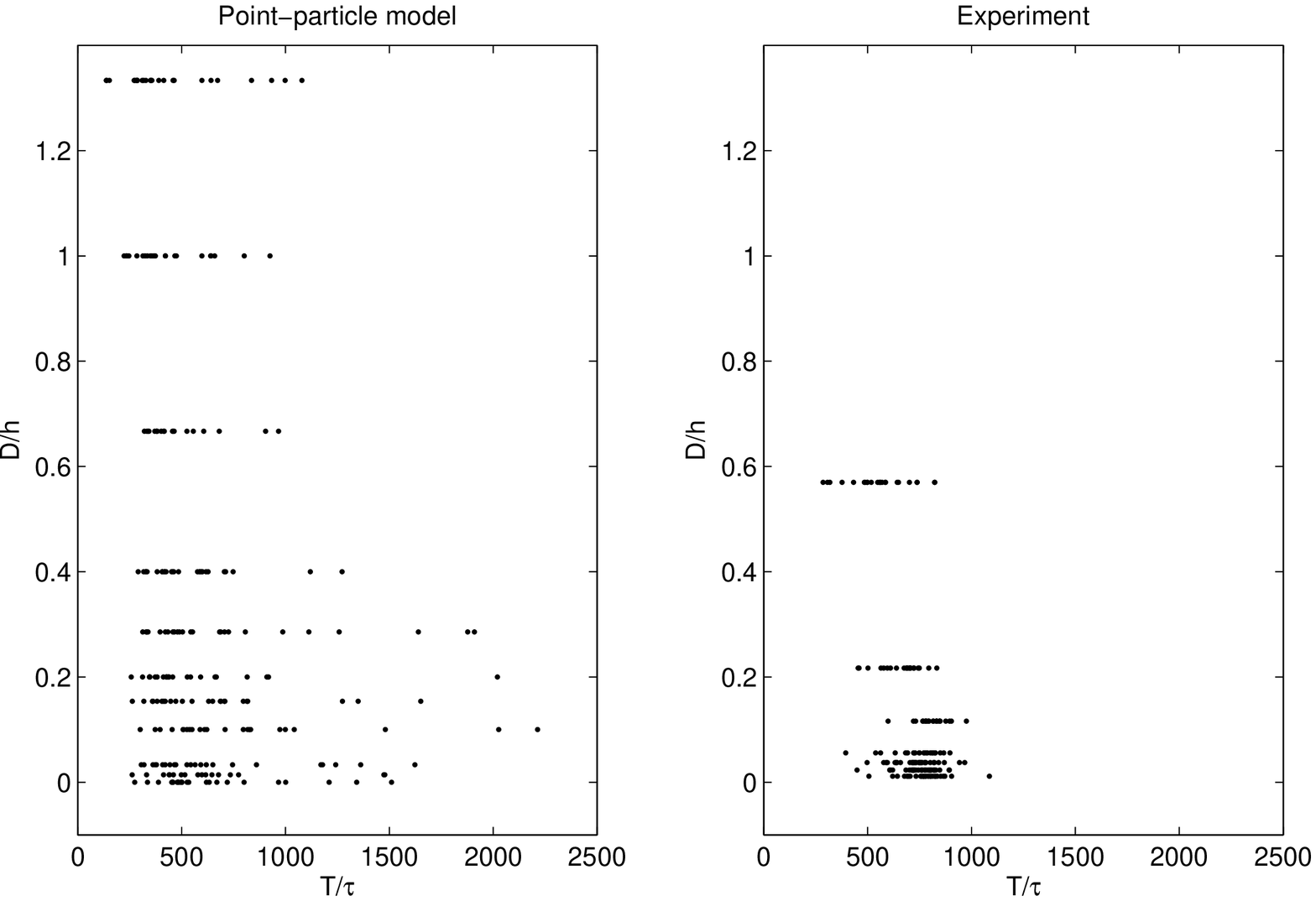}}
\end{center}
\vspace{-0.7cm}
\caption{Destabilization lengths and times of individual drops. Left: the point-particle model; right: experiments.}\label{Tco}
\end{figure*}

The statistics of destabilization lengths and times for individual drops is shown in Fig.~\ref{Tco}. In the experiment and in the model, there exist certain minimal values of both $L$ and $T$ below which the break-up does not occur, because drops need some time to change their shape according to the pattern shown in Fig.~\ref{rys_dla_nn}, in a similar way as it has been observed in point-particle simulations for an unbounded fluid.\cite{art9} 

The destabilization times tend to concentrate around values slightly above this limit, but some of them are significantly larger, even an order of magnitude. 
For the point particles, the range of values attained by both $T$ and $L$ is wider at larger distances $h$ from the wall. No such tendency is observed in the experiments, where in general the spreading of values is smaller than in the simulations, slightly for $L$ and significantly for $T$. The smaller spreading can be explained by the stabilizing effect of the lubrication interaction between close surfaces of the particles.\cite{EJW2006} ''Evaporation'' of particles from the drop by statistical fluctuations is slowed down owing to this effect. In both the experiments and the simulations, the range of the relative differences between destabilization times of individual drops is very large, and it increases when $h$ is getting smaller. The spreading seems to represent chaotic nature of the particle dynamics.\cite{art1}

\begin{figure}[ht]  
\begin{center}
{\includegraphics[width=9.6cm]{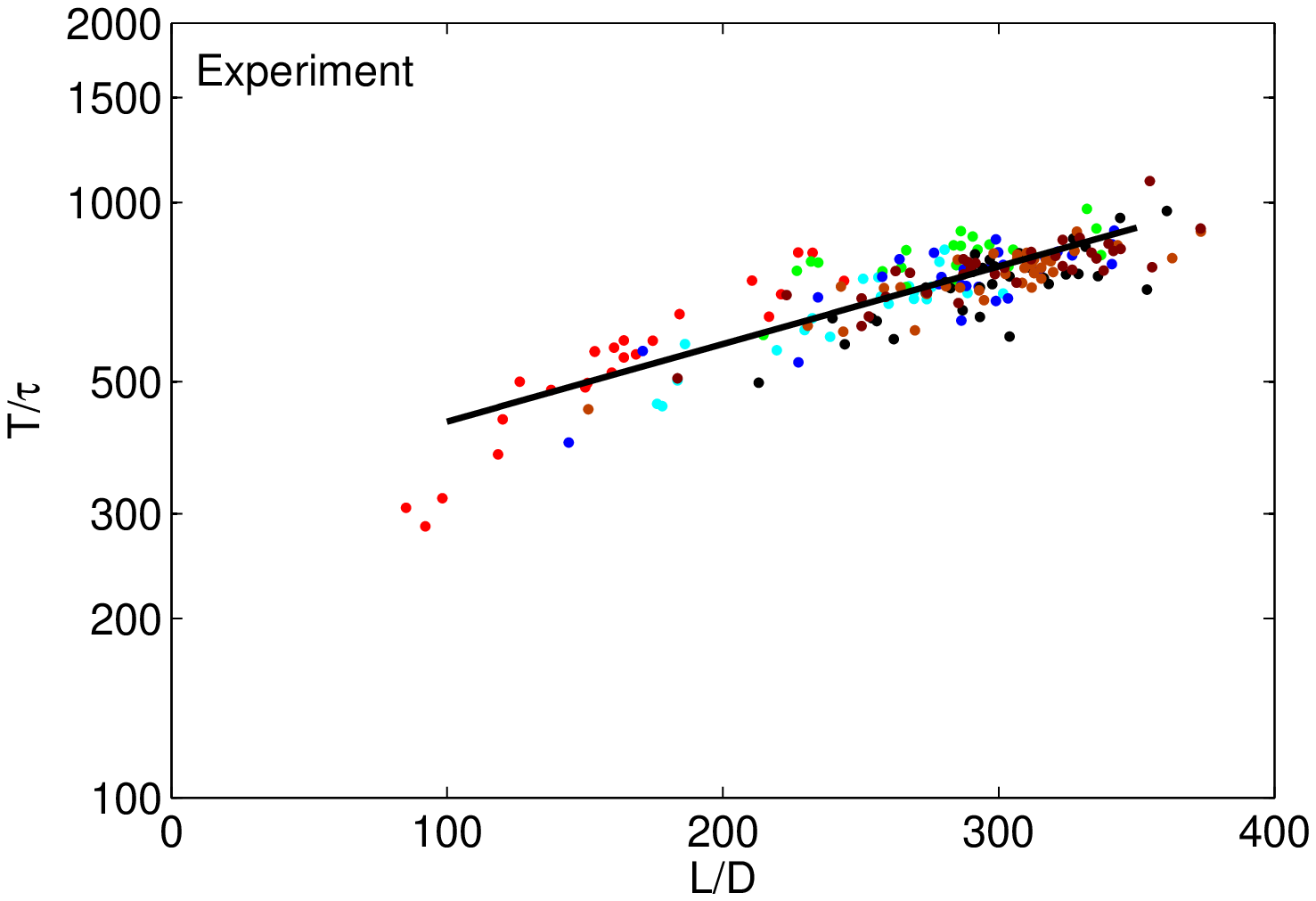}}
{\includegraphics[width=9.6cm]{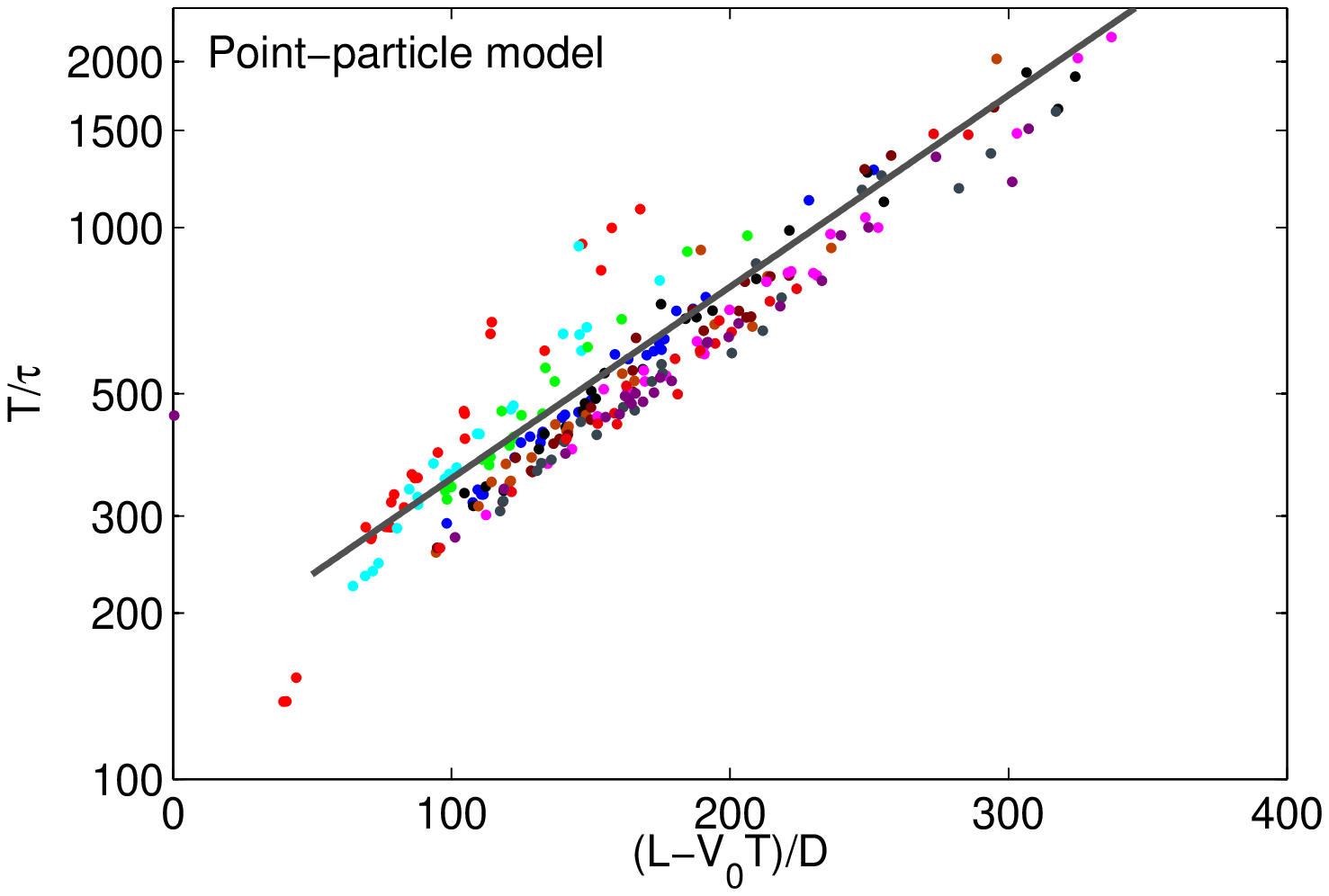}}\\
\end{center}
\vspace{-0.6cm}
\caption{A universal linear scaling of $\ln(T/\tau)$ with $L/D$ for all the destabilization events; top: the  experiment, bottom: the point-particle model. [Color on-line. Top: $h/D$=1.75 (red), 4.63 (cyan), 8.62 (green), 17.9 (blue), 26.8 (black), 43.1 (red-brown), 87.7 (cherry). Bottom:
$h/D$=0.75 (red), 1.0 (cyan), 1.5 (green), 2.5 (blue), 3.5 (black), 5.0 (red-brown), 6.5 (cherry), 10 (magenta), 30 (battleship gray), 70 (dark pink), $\infty$ (purple).] } \label{fig11}
\end{figure}

In Fig.~\ref{fig11}, we investigate the relation between the destabilization time and length, 
%
plotting $T/\tau$ (in a logarithmic scale) versus the destabilization length (in a linear scale) for all individual suspension drops, separately for the experiment (top panel) and the point particle model (bottom panel). In each single panel, $\ln(T/\tau)$ as a function of the destabilization length is well approximated by one linear relation (a least-squares fit) for all the drops. Deviations are observed only for the closest distances from the wall ($h/D=1.75$ for the experiment, red on-line, and $h/D=0.75,1$ for the point-particles, red and cyan on-line). The solid lines in Fig.~\ref{fig11} correspond to the least squares fits $\ln T/\tau = (0.0030 \pm 0.0001) L/D + (5.76 \pm 0.03)$ for the experiments, 
and $\ln T/\tau = (0.0079 \pm 0.0002) (L-V_0T)/D + (5.06 \pm 0.03)$ for the point-particle model.

The experimental slope is less steep than the numerical one, and the constant term is higher. To compare the plots, the point-particle simulations need to be transformed from the frame of reference moving with the Stokes velocity $V_0$ to the laboratory frame. This issue will be discussed at the end of this section.

\begin{figure}[h]  
{\includegraphics[width=9.6cm]{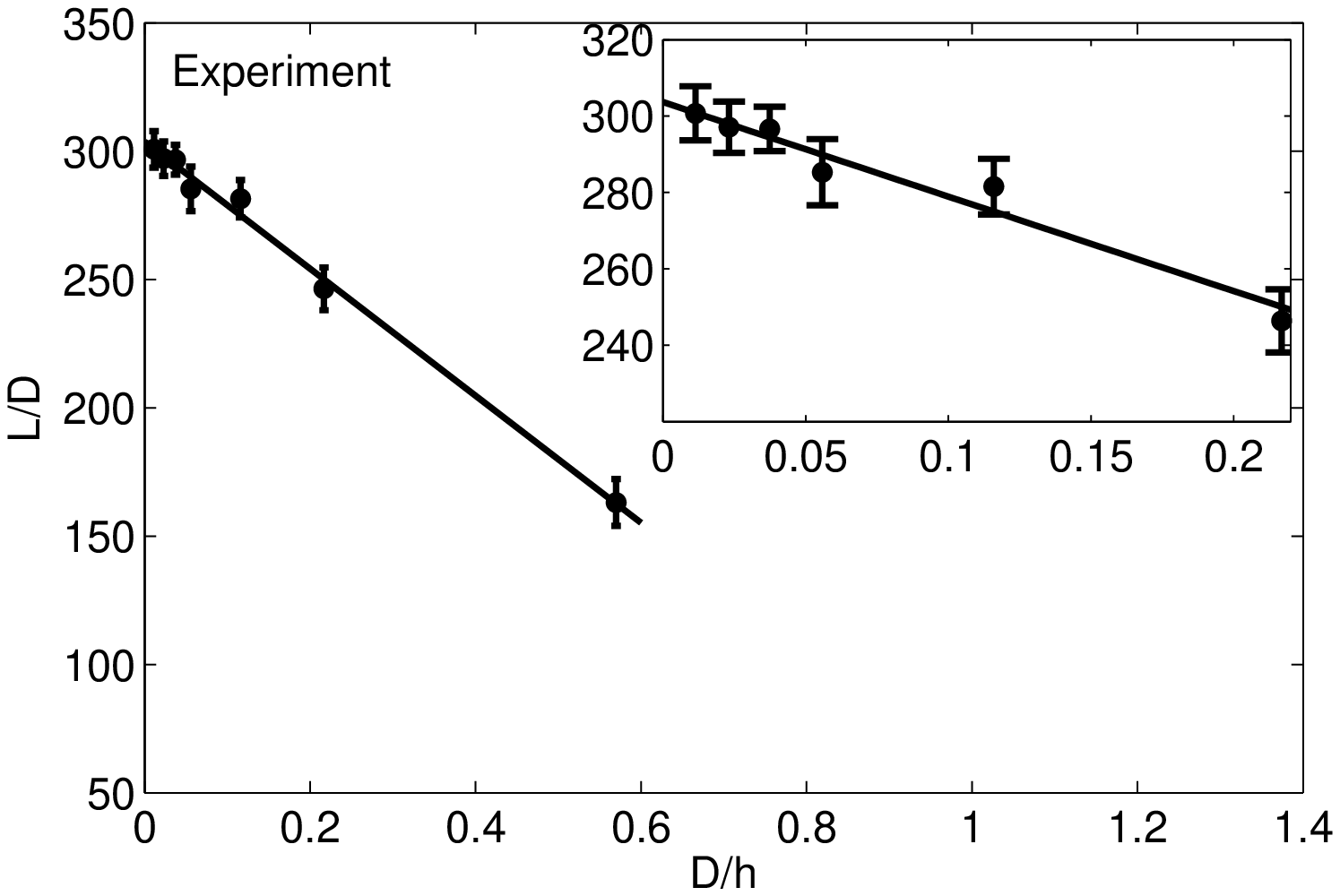}}\\
{\includegraphics[width=9.6cm]{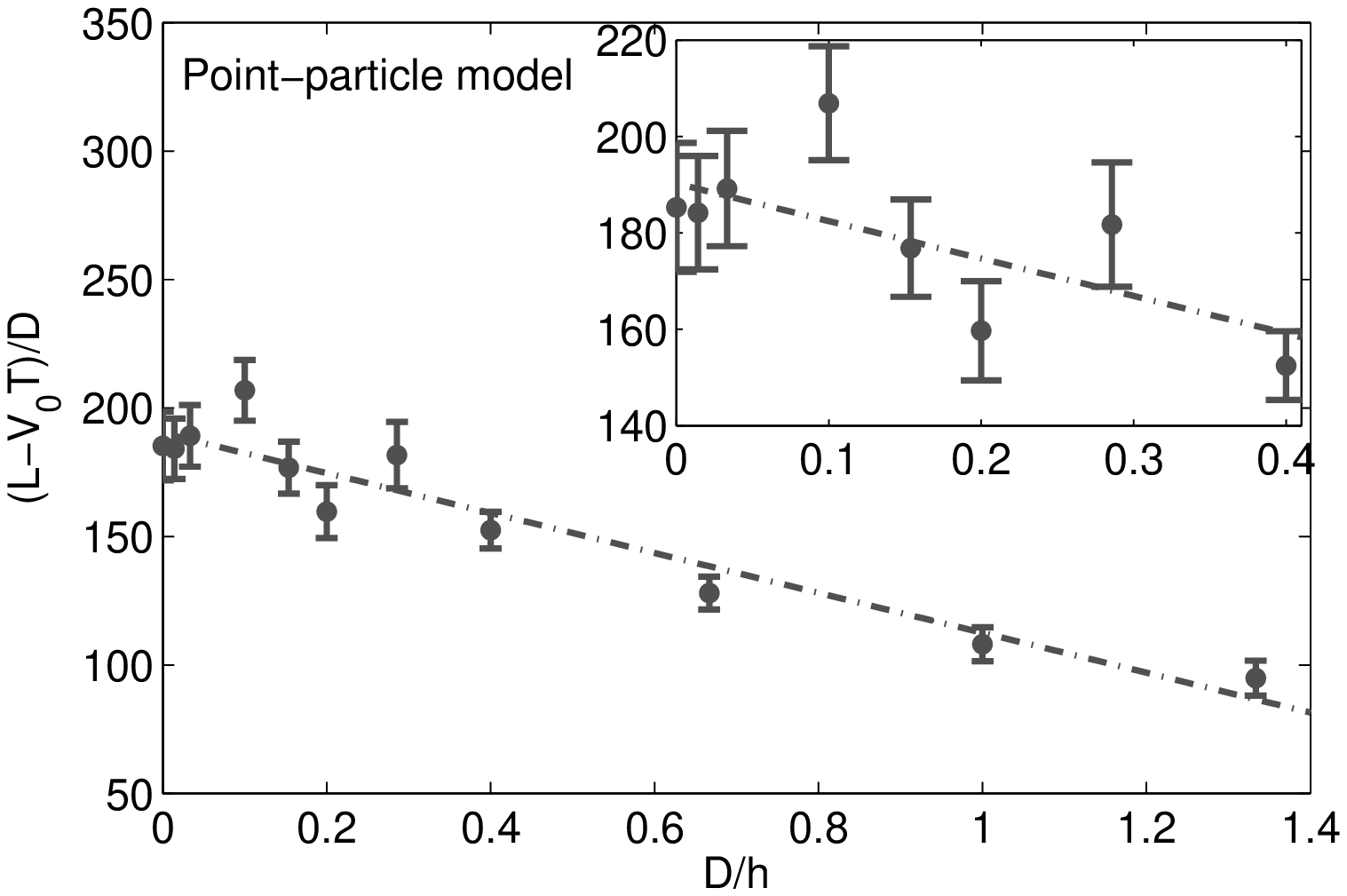}}
\caption{The average destabilization length versus the inverse distance from the wall. Top: experiments, bottom: point-particle model.
}\label{figLD}
\end{figure}

The experimental and numerical average destabilization lengths are compared in Fig.~\ref{figLD}. 
In both cases, $L/D$ is approximately a linear function of the inverse distance from the wall, $D/h$. In detail, the least squares fits in Fig.~\ref{figLD} are $L/D = (-247 \pm 8) D/h + (304 \pm 2)$ (experiment, solid line), and 
$(L-V_0 T)/D = (-78 \pm 9) D/h + (190 \pm 5)$ (point-particles, dash-dotted line). 
For a given $D/h$, the experimental values of $L/D$ decay faster with the decreasing $h$ than the numerical ones. The fast decay of $L/D$ can be related to the increased rate of the particle loss at smaller $h$, observed in the experiments.

\begin{figure}[h]  
{\includegraphics[width=9.6cm]{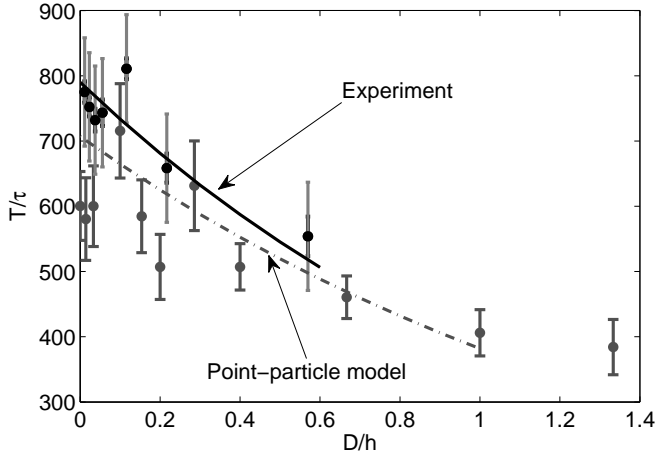}}
\caption{The average destabilization time versus the inverse distance from the wall. Gray: point-particle model, black:  experiments. Dash-dotted and solid lines follow from fits shown in Figs.~\ref{fig11} and \ref{figLD}.} \label{time}
\end{figure}

The average destabilization times, $T/\tau$, are compared in Fig.~\ref{time}. 
The linear fits from Figs.~\ref{fig11} and \ref{figLD} result in the curves plotted in Fig.~\ref{time}. 
The experimental times are slightly larger than the numerical ones, but the difference is smaller than the error bars. We can conclude that the experimental and numerical destabilization times are approximately the same, so that the influence of the drop distance from the wall on the destabilization time found in the experiments is captured well by the point-particle model.

The values of the destabilization times $T/\tau$, listed in Table~\ref{tabIV} and shown in Fig.~\ref{time}, allow to 
estimate the shift, $V_0 T/D=(V_0/2\upsilon) (T/\tau)$, which transforms the point-particle positions at the moment of break-up, from the moving to the laboratory frame of reference. 
For the model of the drop internal structure, described in Sec.~\ref{3.5}, the parameter $V_0/2\upsilon$ varies between 0.01 and 0.03, depending on a specific choice of the particle diameter $d_m$. The shift $V_0 T/D$ to the laboratory frame of reference results in the destabilization length value $L$ increased by around 3\% and 8\%, respectively. This is a small change, and 
in the laboratory frame of reference, there is still a positive difference between the experimental and numerical destabilization lengths. 
This effect will be discussed in the next section.

\section{Conclusions}\label{6}
In this paper, we have analyzed experimentally and numerically the influence of a vertical hard wall on the dynamics of particles settling under gravity in a viscous fluid at low-Reynolds-number. Initially, a large number of close particles was randomly distributed inside a spherical volume. We have shown that the wall significantly affects the dynamics of this group of particles (called a suspension drop). 

The essential effect is that, in the presence of the wall, a drop breaks up faster and it travels a smaller distance as a cohesive entity. It has been shown that the average destabilization time $T$ and distance $L$ are 
smaller for a closer distance $h$ of the drop center from the wall, with approximately linear dependence of $T$ and $L$ on $D/h$, for $h$ larger or comparable to the drop diameter $D$. Otherwise, 
the  typical pattern of the drop evolution is essentially the same with and without a wall, both in the experiments and the numerical simulations, as illustrated in Fig.~\ref{rys_dla_nn}. 

The investigations of this work were challenging, both experimentally and numerically. The measurements in glycerol are difficult, owing to the temperature dependence of the dynamic viscosity, and the tendency of glycerol to capture water from the air. In future investigations, another fluid, e.g. a silicon oil, might be more practical.
 
The numerical results have been evaluated with the use of the point-particle (point-force) model, in the Stokes flow regime. Such a model has been widely applied for an unbounded fluid. We have constructed and applied the point-particle model close to a vertical wall. For identical point forces, this model is very simple, because the particle dynamics does not depend on the particle radius. 

A challenge is also the comparison between the point-particle model and the experiment. It is known from the theoretical foundations and examples studied in the literature~\cite{art27} that the model is too simple to accurately reproduce the drop settling velocity. Therefore, the present paper can be also considered as a case study to estimate the accuracy of the point-particle model. 

For a given distance from the wall, the average destabilization time observed in the experiments and computed in the point-particle simulations is approximately the same. In the experiments, the destabilization length is greater than in the simulations, with the difference not exceeding 30\% far from the walls, and systematically decreasing to zero with decreasing distance between the drop and the wall.

It seems that, in the experiments, the time-dependent drop velocities are higher than in the point-particle simulations, although initially they were equal to each other - faster drops reach a longer distance, if they break at the same time. Drop velocities decrease with time, because drops gradually loose particles. It seems that, in the experiments, a smaller fraction of particles is left behind the drop than in the simulations. In other words, it seems that the experimental drops
tend to keep a larger fraction of particles, measured with respect to the initial number,  and this can be the reason why they move faster than the numerical drops.
This effect can be caused by hydrodynamic interactions between close particle surfaces, which bind them together more strongly than in the case of point-particles. Such a stabilizing effect of lubrication between the particle surfaces has been shown and discussed e.g. in Ref.~[\onlinecite{EJW2006}].

To check this hypothesis, measurements of the number of particles lost in the tail would be worthwhile, in comparison to the results of the point-particle simulations. The evolution of drops, which contain spherical particles, will be the subject of a future numerical study, based on the multipole algorithm  of solving the Stokes equations and the numerical code 
{\sc hydromultipole}.\cite{Cichocki-Ekiel_Jezewska-Wajnryb:1999}

An interesting result is also a statistical distribution of destabilization times and distances. Both $T$ and $L$ were shown to differ significantly from each other, even an order of magnitude, for different random configurations of the particles, owing to the chaotic nature of the particle dynamics. The smaller spreading of values in the experiments can be also explained by a stabilizing effect of the particle surfaces.

The results obtained in this paper can be used in practical applications, such as sedimentation of micro-particles clouds, small milk or ink drops, in micro-channels or close to container walls.

\acknowledgments
The authors acknowledge financial support from the EU COST Action P21 ``Physics of droplets''. A.M. acknowledges support from the Uwe Schaflinger fund through a scholarship.  
The work of A.M. and M.L.E.J. was supported in part by the Polish Ministry of Science and Higher Education grant 45/N-COST/2007/0.
The numerical calculations were done at the Academic Computer Center in Gda\'{n}sk.

\appendix
\section{Estimation of statistical and systematic errors in the experiments}\label{error}
For a sequence of experiments $i=1, ...,n$ with fixed values of $\phi,\;D,$ and $h$, the standard error of the mean, $\bar{ \chi } = \frac{1}{n} \sum_{i=1}^{n} \chi_{i}$, (with $\chi$ standing for $L$ or $T$, and values of $\bar{L}$ and $\bar{T}$ given in Table ~\ref{T1}) was calculated from the standard formula, 
\begin{equation}
S_{\chi}=\left(  \frac{1}{n(n-1)} \sum_{i=1}^{n} \left( \chi_{i} - \bar{\chi} \right)^{2}  \right)^{\frac{1}{2}}.
\end{equation}

During the experiments we observed large variations of the measured quantities from day to day. They might be related to small temperature variations, temperature gradients, convection of the fluid, water at the interface of the glycerol, or other reasons. Such a systematic error was estimated in the following way. First, we computed separately average values measured on a specific day $k=1, ..., m$,
\begin{equation}
 \bar{\chi}_{k} = \frac{1}{i_{k}} \sum_{i=1}^{i_{k}} \chi_{i,k},
\end{equation}
where 
$i=1,...,i_{k}$ labels experiments performed on the day $k$. The corresponding standard deviation,
\begin{equation}
 \sigma_{\chi_{k}} = \sqrt{\frac{1}{i_{k}-1} \sum_{i=1}^{i_{k}}(\chi_{i,k}-\bar{\chi}_{k})^2},
\end{equation}
was then used to determine the weight $w_{k}= 1 /\sigma^2_{\chi_{k}}$ of the measurements performed on the day $k$, and to determine, for a given value of $h$, the average value of $\chi$ and the standard deviation, 
\bee
\bar{\chi}_{w}  &=& \frac{\sum_{k=1}^{m} \chi_{k} w_{k}}{\sum_{k=1}^{m}w_{k}}\\
\sigma _{sy}&=&\left(  \frac{1}{m-1} \sum_{k=1}^{m} \left( \chi_{k} - \bar{\chi}_{w} \right)^{2}  \right)^{\frac{1}{2}},\label{sec:3}
\eee
Finally, we estimated the statistical error as the average of $\sigma _{sy}$ over all such distances $h$, for which measurements were taken on different days.

\end{document}